\documentclass[a4paper,useAMS,usenatbib]{mn2e}
\usepackage[ left=15mm, top=30mm, bottom=30mm, right=15mm]{geometry}

\usepackage{graphicx}
\usepackage{amsmath}
\usepackage{dsfont}

\usepackage{amssymb}
\usepackage{amsmath} 
\usepackage{subfigure} 
\usepackage{epsfig}
\usepackage{appendix}
\usepackage{xspace}
\usepackage{multirow}
\usepackage{mathptmx}
\usepackage{url}
\usepackage{xcolor}
\usepackage{footnote}
\usepackage{natbib}
\usepackage[colorlinks=false,linkcolor=black, citecolor=green, linktoc=page, urlcolor=cyan, 
pdffitwindow=false]{hyperref}
\usepackage{graphicx}
\usepackage{wrapfig}
\usepackage{epstopdf}
\usepackage{acronym}

\def\reff@jnl#1{{\rm#1\/}}
\def\aj{\reff@jnl{AJ}}         
\def\araa{\reff@jnl{ARA\&A}}      
\def\apj{\reff@jnl{ApJ}}        
\def\apjl{\reff@jnl{ApJ}}        
\def\apjs{\reff@jnl{ApJS}}       
\def\aap{\reff@jnl{A\&A}}        
\def\aapr{\reff@jnl{A\&A~Rev.}}     
\def\aaps{\reff@jnl{A\&AS}}       
\def\mnras{\reff@jnl{MNRAS}}      
\def\physrep{\reff@jnl{Physics Reports}}
\def\prd{\reff@jnl{Phys.Rev.D}}     
\def\prl{\reff@jnl{Phys.Rev.Lett}}   
\def\pasp{\reff@jnl{PASP}}       
\def\pasj{\reff@jnl{PASJ}}       
\def\nat{\reff@jnl{Nature}}       
\def\jcap{\reff@jnl{JCAP}}   
\def\memsai{\reff@jnl{MemSAI}} 
\def\na{\reff@jnl{New Astronomy}}       
\def\procspie{\reff@jnl{SPIE}}       
\def\pasa{\reff@jnl{PASA}}       

\usepackage{color}

\def\Sref#1{$\S$\ref{#1}\xspace}

\def\Fref#1{Figure~\ref{#1}\xspace}

\def\Tref#1{Table~\ref{#1}\xspace}

\def\Aref#1{Appendix~\ref{#1}\xspace}
\def\Cref#1{Chapter~\ref{#1}\xspace}

\begin{document}
\twocolumn 

\title{An Integrated System at the Bleien Observatory\\ for Mapping the Galaxy}
\author[C.~Chang, C.~Monstein et al.]{Chihway Chang,$^{1*}$ Christian Monstein,$^{1\dagger}$ Joel Akeret,$^{1}$ Sebastian Seehars,$^{1}$
\newauthor
Alexandre Refregier,$^{1}$ Adam Amara,$^{1}$ Adrian Glauser,$^{1}$ Bruno Stuber$^{2}$ \\
$^{1}$Institute for Astronomy, Department of Physics, ETH Zurich, Wolfgang-Pauli-Strasse 27, 8093 Z\"urich, Switzerland \\
$^{2}$University of Applied Science, FHNW, Brugg, Switzerland \\
$^{*}$chihway.chang@phys.ethz.ch \\
$^{\dagger}$monstein@astro.phys.ethz.ch}

\date\today
\maketitle

\begin{abstract}

We describe the design and performance of the hardware system at the Bleien Observatory.  
The system is designed to deliver a map of the Galaxy for studying the foreground contamination of 
low-redshift (z=0.13--0.43) H$_{\rm I}$ intensity mapping experiments as well as other astronomical 
Galactic studies. This hardware system is composed of a 7m parabolic dish, a dual-polarization corrugated 
horn feed, a pseudo correlation receiver, 
a Fast Fourier Transform spectrometer, and an integrated control system that controls and monitors the 
progress of the data collection. The main innovative designs in the hardware are (1) the pseudo 
correlation receiver and the cold reference source within (2) the high dynamic range, high frequency 
resolution spectrometer and (3) the phase-switch implementation of the system. This is the first time 
these technologies are used together for a L-band radio telescope to achieve an electronically stable 
system, which is an essential first step for wide-field cosmological measurements. This work demonstrates 
the prospects and challenges for future H$_{\rm I}$ intensity mapping experiments.  \\

\end{abstract}


\section{Introduction}
In the coming decades, a wealth of astronomical data in radio wavelength will become 
available through large survey projects and telescopes such as 
LOFAR\footnote{\url{ http://www.lofar.org}} \citep{vanHaarlem2013}, 
GMRT\footnote{\url{http://www.ncra.tifr.res.in/ncra/gmrt}} \citep{Paciga2013},
PAPER\footnote{\url{ http://eor.berkeley.edu}} \citep{Ali2015}, 
CHIME \citep{Bandura2014}, 
BINGO \citep{Battye2012, Battye2013}, 
HERA \citep{Pober2014}, Tianlai \citep{Chen2012},
and SKA\footnote{\url{ http://www.skatelescope.org}} \citep{Mellema2015}. 
Most of these projects aim at measuring signal from the redshifted 21 cm H$_{\rm I}$ emission line, 
from either low-redshift large-scale structure \citep[e.g.,][]{Battye2012, Masui2013} or high-redshift 
Epoch of Reionization \citep[e.g.,][]{Furlanetto2006, Paciga2013}. Many of these surveys will suffer 
from foreground contamination of our own galaxy, the Milky Way, which is typically several 
orders-of-magnitude larger than the cosmological signals of interest. As a result, a good 
understanding for this foreground component is crucial for the interpretation of the cosmological 
measurements of interest. This need for a deeper understanding of the Milky Way at the 
relevant H$_{\rm I}$ wavelength is the main driver for this work.

To date, only a few of wide-field Galactic maps exist and are available for the community to use 
to study the H$_{\rm I}$ foreground \citep[see][and reference therein]{deOliveira-Costa2008}. 
Furthermore, the frequency coverage has been fairly sparse across the different maps. The most 
commonly used map around 1 GHz has been the so-called ``Haslam map'' at 408 MHz 
\citep{Haslam1982}. This map has been re-processed several times \citep{Remazeilles2015}, but 
almost no new data has been taken systematically since then.
The main reason for the lack of wide-field maps at these frequencies is that historically, this has 
not been a wavelength window for cosmological measurements. In addition, the generation of these 
maps require significant observing time even at moderate resolution. With the next generation of 
H$_{\rm I}$ cosmological experiments being built, foreground maps at the radio L-band are 
becoming more important.

\begin{figure}
  \begin{center}
   \includegraphics[scale=0.32]{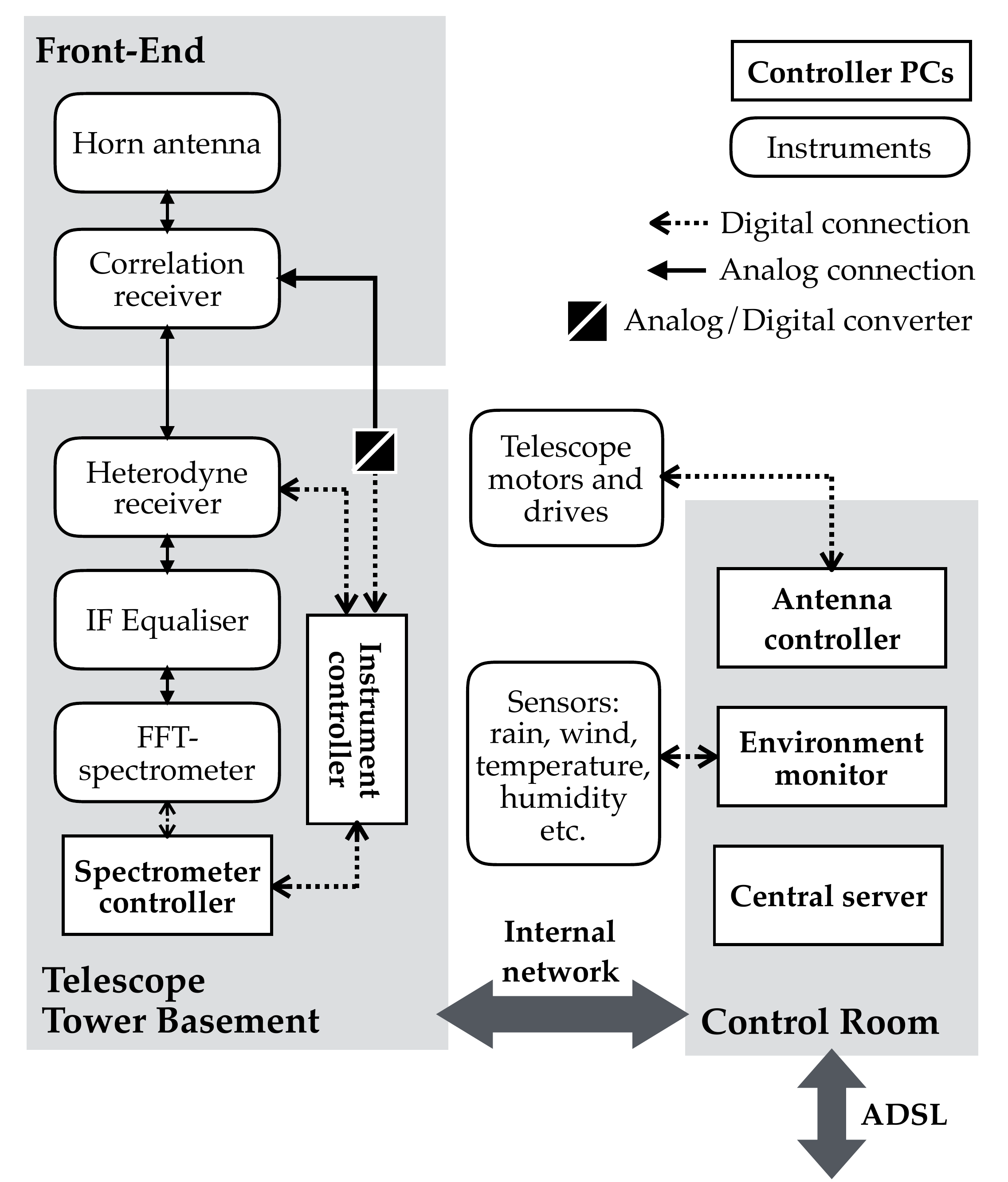}  
  \caption{Schematic illustration of the entire system for this project. The three grey 
  boxes indicate the three different locations where the instrumentation are -- the front-end 
  unit mounted inside the telescope dish, the telescope tower basement, and the control room located 
  about 60 m away from the telescope. The white squares indicate controller PCs whereas the white 
  rounded squares indicate instruments. All computers in the control room connect to each other 
  as well as the external network. }
  \label{fig:overview}
  \end{center}
\end{figure}

In this work, we investigate the possibility of a simple system that could be used 
for these foreground maps. This system consists of a dedicated single-dish 
telescope that scans the sky over several months with very low human intervention 
during the observation.
The hardware system is set up at the 7m telescope at the Bleien Observatory, 
operating in the frequency range 
990--1260 MHz. We emphasise the innovative yet low-cost design that resulted in significant 
improvement in data quality compared to the more conventional systems. This work continues 
from \citet{Chang2015} as a series of studies at the Bleien Observatory in preparation for future 
H$_{\rm I}$ intensity mapping experiments.

The paper is organized as follows. In \Sref{sec:system}, we describe our overall system at the Bleien 
Observatory, both the hardware and the controlling/monitoring mechanism. In \Sref{sec:commission}, we 
evaluate the performance of the system and the improvement over the previous system with both 
laboratory measurements and on-sky measurements. Our conclusions are summarized in 
\Sref{sec:conclusion} and additional tests of the hardware system are presented in the Appendix. 
A companion paper (Akeret et al. in prep, hereafter A16) describes the analysis software and some 
early processed data. 

\section{System description}
\label{sec:system}
In this section, we describe the instrument and software chain from the horn antenna where the signal 
arrives to the pre-processing step of the data which feeds into the science analysis. \Fref{fig:overview} 
shows a schematic illustration of the full system. \Tref{tab:instrumentspecs} sums up some 
of the main characteristics of the system, some of which were characterised during the 
commissioning period.  

\begin{figure}
  \begin{center}
   \includegraphics[scale=0.8]{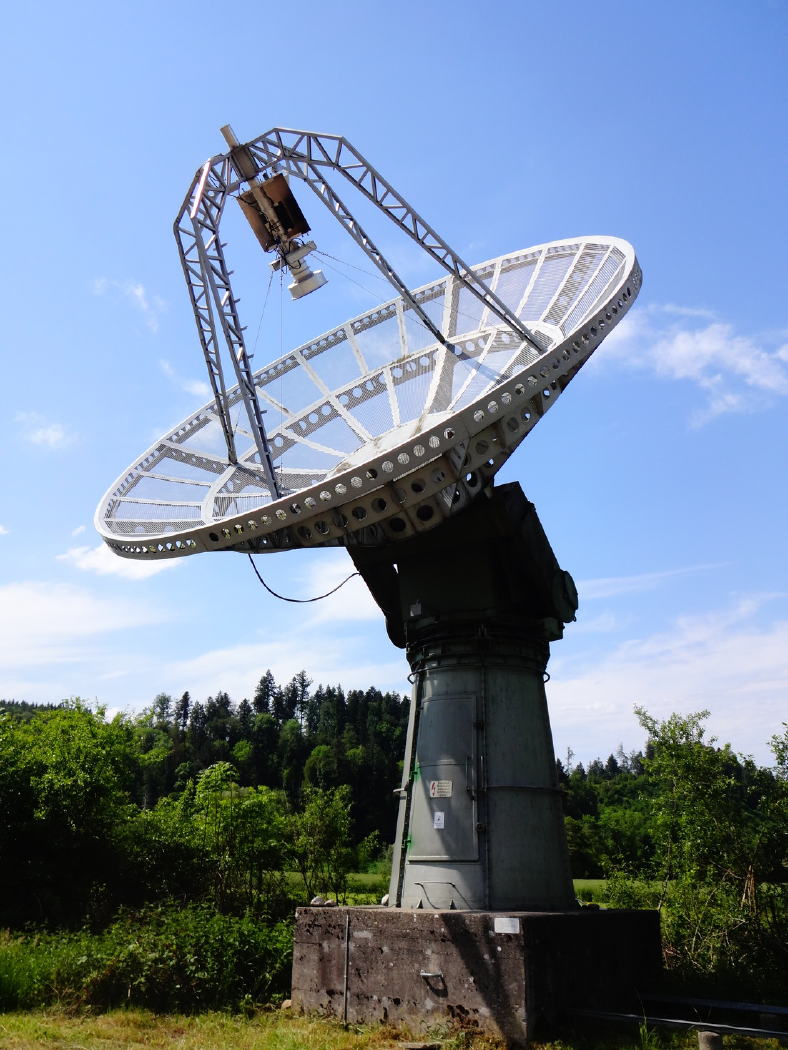}  
  \caption{The 7m telescope at the Bleien Observatory and the new feed horn. From this image, 
  one can see the telescope tower, the 7m parabolic dish, and the horn antenna mounted at the 
  focal plane.}
  \label{fig:dish}
  \end{center}
\end{figure}

\subsection{The Bleien Observatory}
\label{sec:bleien}

The Bleien Observatory\footnote{\url{http://www.astro.ethz.ch/research/Facilities/Radioteleskop_Bleien}} 
is composed of two parabolic dishes of 5m and 7m diameter and a control room housing supporting 
instruments. The observatory is located at Bleien, Switzerland, about 50 km south-west of Zurich 
(geographic latitude $47^{\circ}~20'~23''$ north, longitude $8^{\circ}~6'~42''$ east, and altitude 
469 m). The observatory was built in 1979, which, at the time was a nearly RFI-free location ideal for 
radio observations. Today, however, the RFI situation is worsened due to the ever-increasing TV and 
radio stations as well as mobile phone transmitters. Although the site is protected within a 1.5 km via 
the federal office of communication (OFCOM), significant RFI contamination can still be seen in our 
data, which is one of the main challenges for the data reduction as we will discuss below.

The 7m dish (F/D 0.507) is used in this work. 
This telescope is composed of a concrete basement, a steel tower, and an old radar 7 m dish produced 
during the second World War. Major upgrades during the years of 2002 -- 2008 allow more precise pointing and 
remote control of the telescope. As the original science objective of the telescope has been solar 
observations, hardware upgrades had to be made to meet the different requirements for cosmological 
H$_{\rm I}$ measurements. The major upgrade on the telescope involves the manufacturing and installation 
of a new dual-polarization, corrugated horn feed described below. \Fref{fig:dish} shows the new horn installed 
at the 7m dish focal plane. 

\subsection{The horn feed}
\label{sec:horn}

The horn is manufactured by the company ANTERAL in Spain. It is designed to optimally reduce 
potential side lobes and ground spill-over. It contains two polarizations with a cross-talk 
below -24 dB. There is a 4-port ortho-mode transducer in the back of the horn of size 
560$\times$560 mm, and the horn itself has a diameter of 450 mm with length 480 mm. The 
horn weighs 22 kg and has an aperture efficiency $>0.6$. It under-illuminates the dish by a factor 
of 0.5, which is designed to minimize the side-lobe pickup from the ground and other RFI sources.

\begin{figure*}
  \begin{center}
   \includegraphics[scale=0.4]{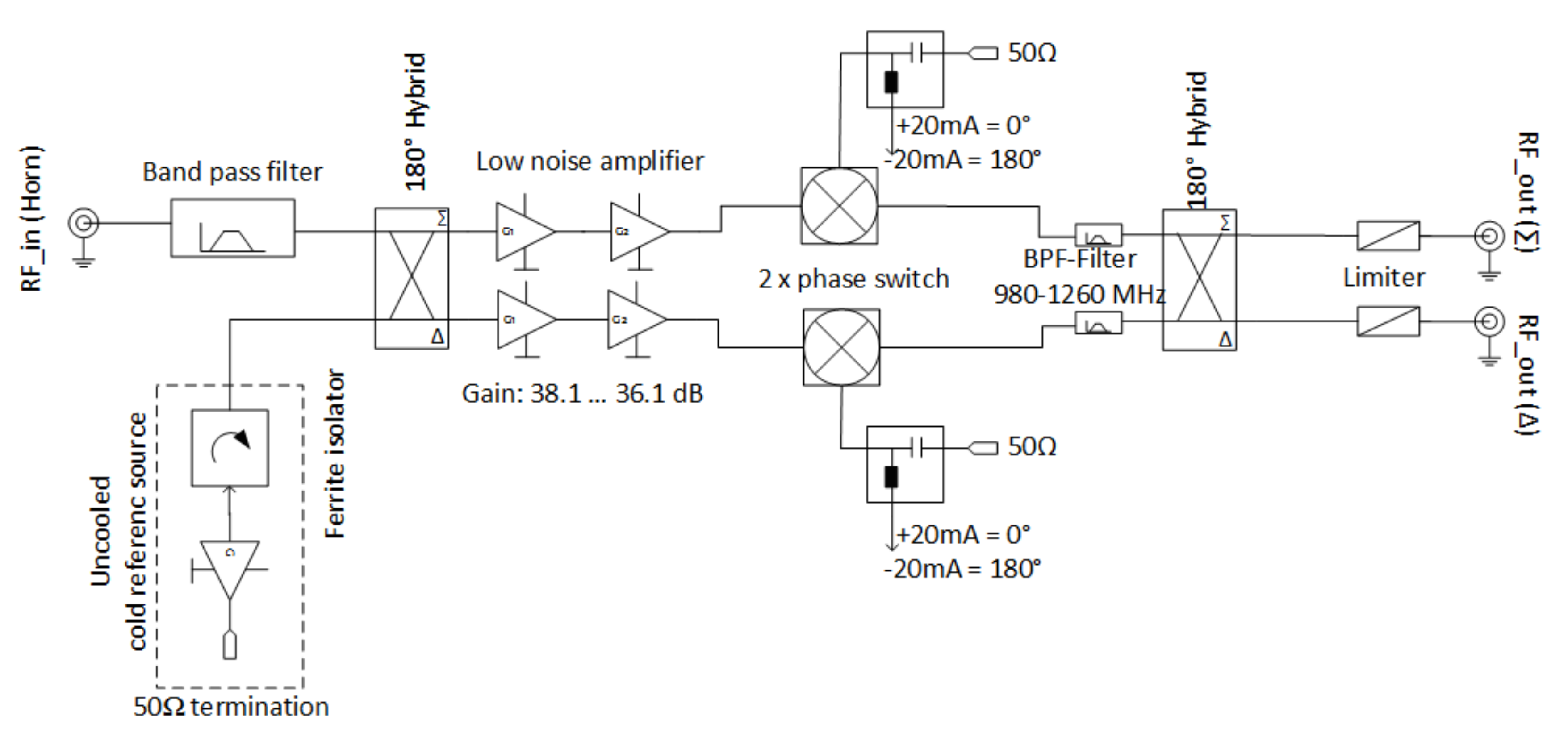}  
  \caption{Schematic diagram of the PCR. Two parallel and almost identical chains are shown, with 
  the electronic signal propagating from left to right. The upper chain is connected to the telescope 
  horn receiver on the left and the lower chain connected to the cold reference source as labeled. 
  The final output of upper chain contains is fed to the heterodyne receiver (\Sref{sec:hrx}), while the 
  lower chain is not used.}
  \label{fig:pcr}
  \end{center}
\end{figure*}

\subsection{The pseudo correlation receiver and the cold reference source}
\label{sec:pcr}

The pseudo correlation receiver \citep[PCR,][]{Mennella2003} is the core subsystem in the whole 
chain of the instrument. Its main function is to reduce the $1/f$-noise\footnote{$1/f$-noise refers 
to the noise component which has a power spectrum that scales with inverse frequency. This means 
that the noise is more correlated over long time-scales and less correlated over short time-scales. 
$1/f$-noise is typically generated in all electronic devices.} generated in the high-gain, 
low-noise amplifiers. The term `correlation receiver' refers to the fact that a reference 
signal is taken together with the sky signal in order to reduce $1/f$-noise introduced by the electronics. 
The term `pseudo' suggests that the correlation operation is implemented via a `sum' operation 
rather than the more conventional `multiplication' and only requires one analog channel per 
polarization. 

A schematic of the electronic layout of the PCR in our system is shown in \Fref{fig:pcr}. 
Two parallel and almost identical chains are used with one chain connected to the telescope horn 
receiver at input and the other connected to a cold reference source described below. The signals 
from both chains are combined to give the ``sum'' signal at the output of the upper chain, which feeds 
to the heterodyne receiver described in the next section. The output of the lower chain contains 
redundant information and is discarded. All components in the PCR are thermally closely linked 
together and controlled to be at a temperature of $18.0^{\circ}\pm0.2^{\circ}$. 

In addition to the conventional PCR design, the voltage phase of the PCR in our system can switch 
between $0^{\circ}$ and $180^{\circ}$. Subtracting the received signal from the two phases allows us 
to further reduce $1/f$-noise introduced by the digital backends since the instrumental $1/f$-noise 
from the two electronic chains are not correlated (while the signal from the sky is). In practice, the phase 
switch is executed every 3 seconds\footnote{The choice of this integration time has been originally 
chosen to minimize the fraction of deadtime between integrations. In the current system, this 
deadtime is no longer a problem, but we keep the integration time as the gain in going to shorter 
integration time is not significant. This also gives a reasonable data volume ($\sim17$MB per 15 
minutes) to handle in the later analyses.}. 
Phase-switch instruments and other beam-switch or frequency-switch instruments that utilize the same 
principle have been used frequently in microwave receivers \citep{Kraus1965, Mennella2003}. However, 
they are not as often used in radio instruments since the long-term baseline stability has not been a 
major requirement for traditional radio observations.  

The cold reference source is designed to provide a similar spectrum at the same noise level
as the science horn to minimize balancing issues in the correlation receiver. Conventionally, 
such a reference source is either implemented through a separate antenna looking to a dedicated
reference sky position, or a resistor cooled with liquid nitrogen. In this work, we demonstrate a 
low-cost alternative of using an electronic source, based on commercially available components: 
the reference source is composed of an inversely connected low-noise amplifier and a broad-band 
ferrite isolator. The additional advantage of an electronic reference is that it does not generate 
undesired interference with the environment and the science signal. The basic principles behind 
this design was first developed in \citep{Frater1981}, but so far there are only few existing 
implementations shown in \citet{Fabry2009, Straub2010, Scheeler2013}.

The remaining elements in the electronic chain shown in \Fref{fig:pcr} are tested in the 
lab as well as in simulations to achieve the ultimate high-gain, low-noise required between the 
antenna and the rest of the chain. We also note that the band-pass filter (BPF) is applied at the 
end of the electronic chain to narrow down the frequency range to close to our final target.

\subsection{The heterodyne receiver}
\label{sec:hrx}
\begin{figure*}
  \begin{center}
   \includegraphics[scale=0.6]{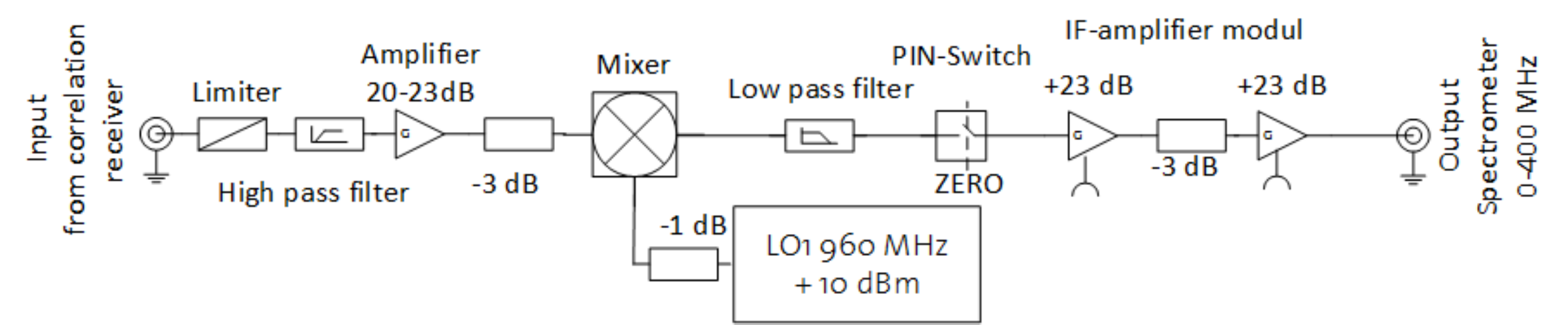}  
  \caption{Schematic diagram of the HRX.}
  \label{fig:hrx}
  \end{center}
\end{figure*}

The heterodyne receiver (HRX) is introduced to resolve the mismatch between the frequency 
range of our expected signal (990--1260 MHz) and the frequency range where the FFT-spectrometer 
operates (below 800 MHz). The HRX down-converts the frequency range of the incoming signals 
using a local oscillator and a mixer circuit. In addition, the HRX provides a gain of $\sim$40 dB, 
which is needed to drive the analog digital converter of the FFT-spectrometer. The HRX also 
includes a high-pass filter which suppresses radiation from nearby mobile phone transmitters at 
$\sim$950 MHz and a low-pass filter to avoid aliasing in the FFT-spectrometer\footnote{The 
bandwidth of the signal has to be limited to half the sampling frequency which in our case is 
1600 MHz.}. The HRX is physically part of the instrument controller to keep cabling short and 
temperature as constant as possible. A schematic diagram of the electronic layout of the HRX 
is shown in \Fref{fig:hrx}. 

\subsection{The FFT-spectrometer}
\label{sec:fft}
\begin{figure}
  \begin{center}
   \includegraphics[scale=0.32]{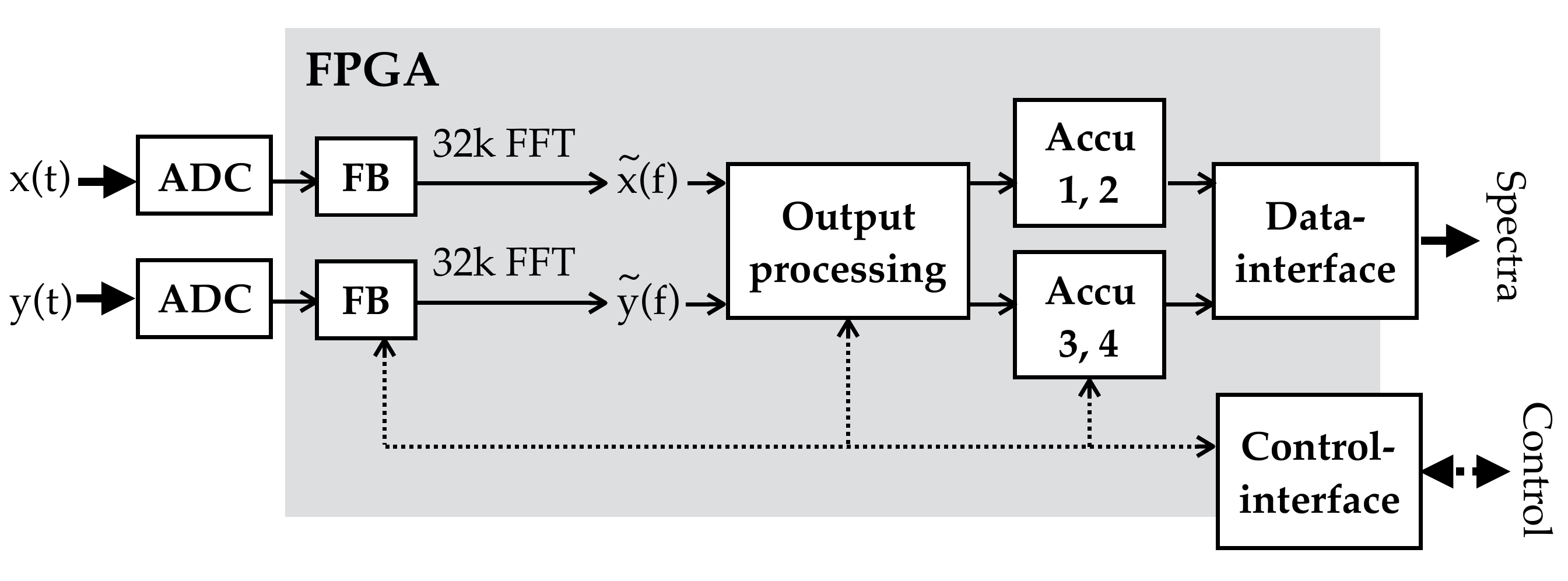}  
  \caption{Schematic diagram of the FPGA unit in the FFT-spectrometer. $x(t)$ and $y(t)$ are the 
  signal from the two polarizations (which we only use one in our case). The analog-to-digital 
  converter (ADC) digitalizes the signal 
  before entering the FPGA unit. After a Fast Fourier Transform (FFT), the signal is processed and 
  accumulated via the accumulator (Accu) for both the spectral power density (P) and the variance 
  of P. We control the FPGA via the control-interface and access the final spectral data output via 
  the data-interface.}
  \label{fig:filterbank}
  \end{center}
\end{figure}

The FFT-spectrometer receives the analog signals output from the HRX after the 
equalizer\footnote{Due to the 25m-long coaxial cables between the PCR and the HRX, there is a 
large spectral slope in the order of 10dB in our frequency range, which is unideal in terms of the 
dynamic range of the spectrometer. The spectrum equalizer is added to attenuate
lower frequencies more than higher frequencies, resulting in a flat final spectrum.}. The signal is first 
digitized and then processed in a Field Programmable Gate Array (FPGA) unit. A schematic 
of the FPGA unit is shown in \Fref{fig:filterbank}. The spectral analysis is based on FFT on a 
32k$\times$32k grid, which in our setting results in a frequency channel separation of 48.8 kHz. 
The main improvement of this FFT-spectrometer over older implementations \citep{Benz2005} 
is its digital filterbank, which convolves the input signal with a chosen filter to better separate the 
channels and reduce side-lobe contamination. 
The built-in filterbank of our FFT-spectrometer includes different windowing functions such as 
Hamming, Hanning, Blackman-Harris, and Flattop. Based on simulations and laboratory tests, 
we find that the Flattop window function results in the best performance in our case. 
The FFT-spectrometer also has 
a high dynamic range of 70dB, which implies that the high-level RFI is less likely to saturate 
and leak into neighbouring bands. Although not used 
directly in our main analysis, the filterbank core has several other powerful functionalities: computing 
sum and difference of the spectra, sideband separation, cross power spectrum, and recording square 
of the power spectrum for kurtosis analysis \citep{Nita2010}. 

As discussed in \Sref{sec:pcr}, we operate the PCR in phase-switch mode. In the default setting 
of the FFT-spectrometer, this means that a file is written every time the phase-switch happens, 
i.e. every $\sim$3 seconds. This is not ideal since it generates a large number of 
single-spectrum files and some latency dead-time in the communication between the 
FFT-spectrometer and the PCR especially in the early stage of the experiment. Nevertheless, given 
the data taken, we describe in \Sref{sec:preprocess} how we process the data into more user-friendly 
formats. Future models of this FFT-spectrometer will improve on this feature. 

\subsection{Instrument control and monitoring}

The full system is controlled via several control PCs inside the telescope tower and the 
control room. The five main PCs are marked as rectangles in \Fref{fig:overview}. They 
communicate with the antenna, the PCR, the HRX, the FFT-spectrometer and other monitoring 
sensors around the observatory. The interfaces are based on the ARDUINO UNO 
micro-controller\footnote{\url{https://www.arduino.cc/en/Main/ArduinoBoardUno}}. 
There is no digital communication between the controller computer and the front-end (including 
the horn antenna and the PCR) to avoid self-produced interference. All information from and 
to the front-end is fed with analog wiring. 

The antenna controller PC is connected to the 7m dish antenna via a commercial dual
drive system (one axis for the azimuth direction and a second axis for the elevation direction).
It is also linked to the internal network and to the Internet for remote access. 
The main functions of this controller PC is to send commands related to the pointing position 
of the telescope. This includes fixed celestial positions, tracking of astronomical sources, and 
specific scanning patterns for special calibration purposes (e.g. 2D raster-scans for pointing 
calibration). A pre-designed survey schedule can be provided to automate long observation 
plans. The control system records the time and position of the telescope, which is measured by 
optical encoders, directly mounted to the axis of the telescope. The positioning resolution is given 
by the number bits. In our case, 12 bits leads to a pointing resolution of $0.088^{\circ}$.

The instrument controller PC communicates with the PCR and the HRX via USB. 
The main tasks of this PC include commanding the phase switch to the PCR and regulating 
the temperature of the high-frequency components in the focal plane unit. Large changes in 
temperature are regulated by a 2-point regulator to heat/cool with Peltier elements, while small 
temperature changes are controlled via Proportional plus Integral plus Derivative (PID) elements. 
Measurement of the ambient temperature, electronics temperature, humidity and the voltage 
supply of the different parts are sent back to the controller computer and recorded for real-time 
monitoring. It also connects to the spectrometer controller PC that controls the FFT-spectrometer.

The environment monitor PC is connected to the internal network and collects information from the 
different environmental sensors in and around the observatory, including wind, rain, temperature 
and humidity. And finally, the central server serves as a dedicated data server and stores all 
the raw spectral, environmental and instrumental data. It also serves as the web server for the 
monitoring website. 
A real-time internal monitoring website is set up for easy monitoring of all the data collection and 
instrument/environment status.

\subsection{Data pre-processing}
\label{sec:preprocess}

As mentioned in \Sref{sec:fft}, the FFT-spectrometer writes out one file per measurement for 
every integration period. In the current setup the system measures a spectral power density 
($P$) and its variance ($P^2$) in both voltage phases. The accumulation period is 
defined to be 146484 (number of samples) $\times$ 20.48$\mu {\rm sec}$ (sampling time) 
$\approx 3.0$ sec. In a continuous operation of several weeks, 
the number of created files rapidly exceeds the capabilities of regular file systems. To 
circumvent this an independent process collects and aggregates the single spectrum files 
in a fixed interval (currently set to be 15 minutes). Additionally, a `time axis' is created that 
hold information about the time when each spectrum was measured. During this aggregation 
process, frequency channels which are outside the band-pass filter are discarded to reduce the 
data volume. Finally, the aggregated data is written to disk in HDF5 file format and a moderate, 
lossless gzip compression is applied. By discarding obsolete data and the file compression the 
data volume is reduced by a factor $3.5$ on average. See A16 for a detailed 
description of the data structure. 

\section{Commissioning}
\label{sec:commission}
A two-stage commissioning process was carried out from mid-November to mid-December 
of 2015. The first stage involves testing the new hardware in the laboratory, while the second 
stage includes installing the system at the observatory and performing on-sky measurements. 
Results from the two stages are presented below. 

For the purpose of illustration, in the following we often compare the performance of our new 
phase-switch correlation system with another system that does not include some of these major 
improvements described in \Sref{sec:system}. 
A convenient choice for this system is the data collected through the other 
polarization of the horn feed. This data will come from exactly the same part of the sky and be 
contaminate by almost the same RFI signals. The signal is then connected to an electronic 
chain that does not have the PCR and phase-switch implementation, and eventually to an 
identical FFT-spectrometer. Effectively, this comparison shows the improvement of our new 
design over a more conventional radio telescope system. When applicable, we refer to this 
other system as the `total-power chain', as there is no phase-switch operation in this chain. 
The schematic layout of the total-power chain would be very similar to \Fref{fig:overview}, 
except that it does not contain the correlation receiver. 

In \Aref{sec:sv_tests}, we present several additional tests that are not directly associated with the 
new electronic system described earlier, but are generally important in terms of the planning 
of a Galactic survey. These tests include the site RFI characterization, the telescope pointing 
accuracy, the beam size, gain and system temperature. The tests were performed both during 
the commissioning period and the ``Science Verification'' stage that followed shortly after 
commissioning (see \Aref{sec:sv} for a brief description of the Science Verification 
observations). 

\begin{figure}
  \begin{center}
   \includegraphics[scale=0.5]{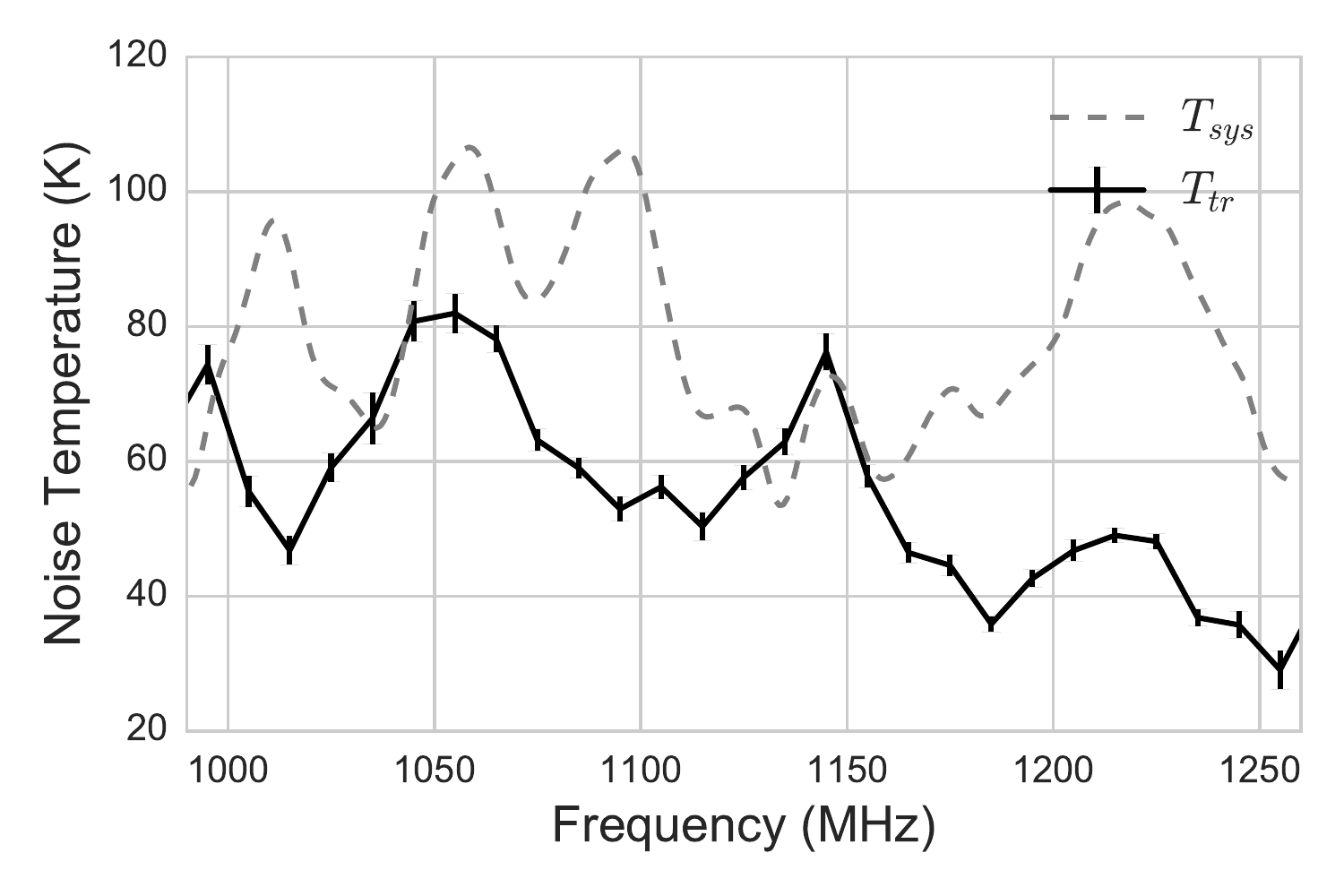}  
  \caption{Mean and uncertainty of the noise temperature of the PCR ($T_{rx}$) measured over 
  10 MHz bins in the frequency range of interest. The black solid line is the average $T_{rx}$ and 
  the grey dashed line shows the total system noise temperature $T_{sys}$.}
  \label{fig:trx}
  \end{center}
\end{figure}

\begin{figure*}
  \begin{center}
   \includegraphics[width=0.45\linewidth]{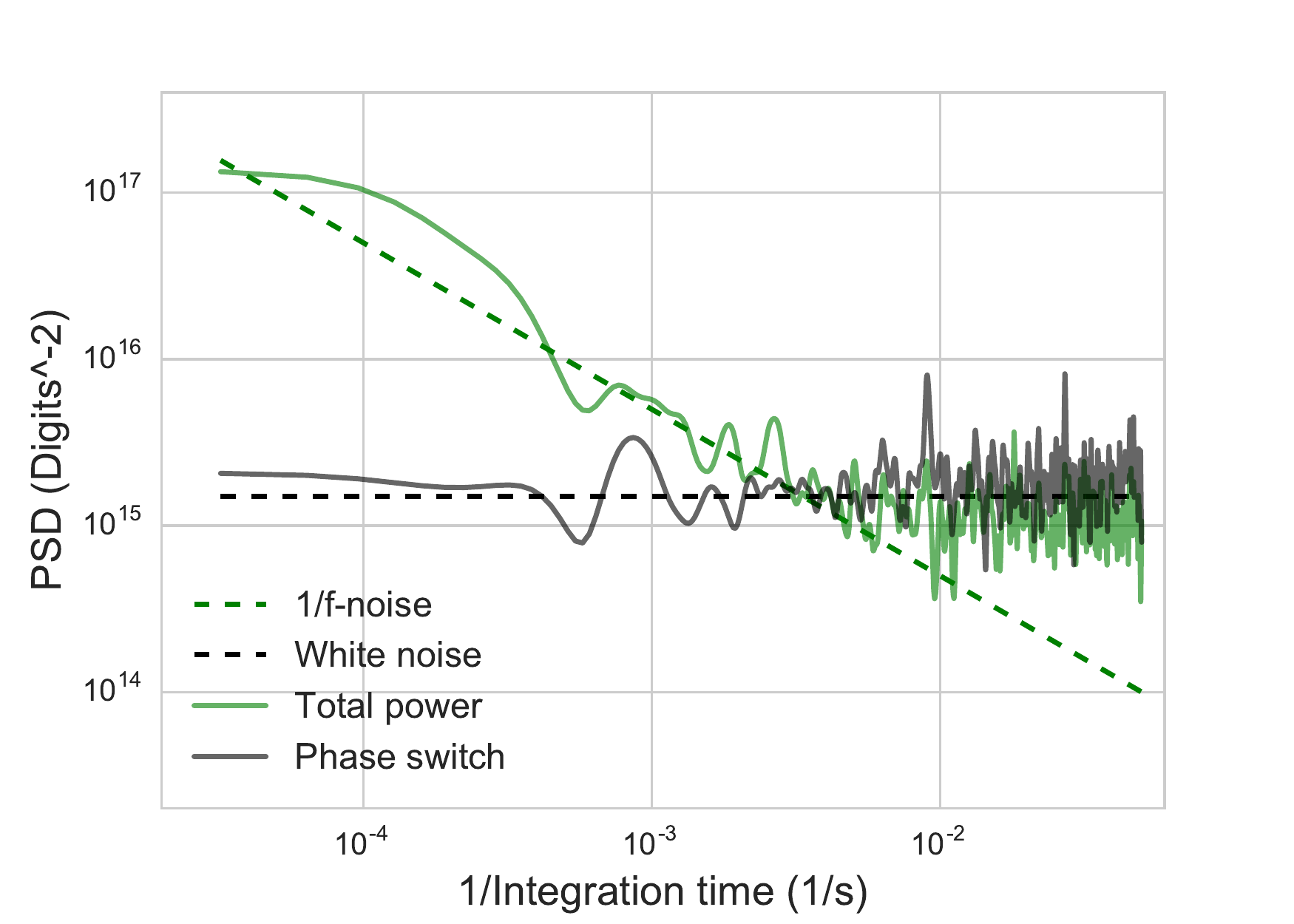}    
    \includegraphics[width=0.45\linewidth]{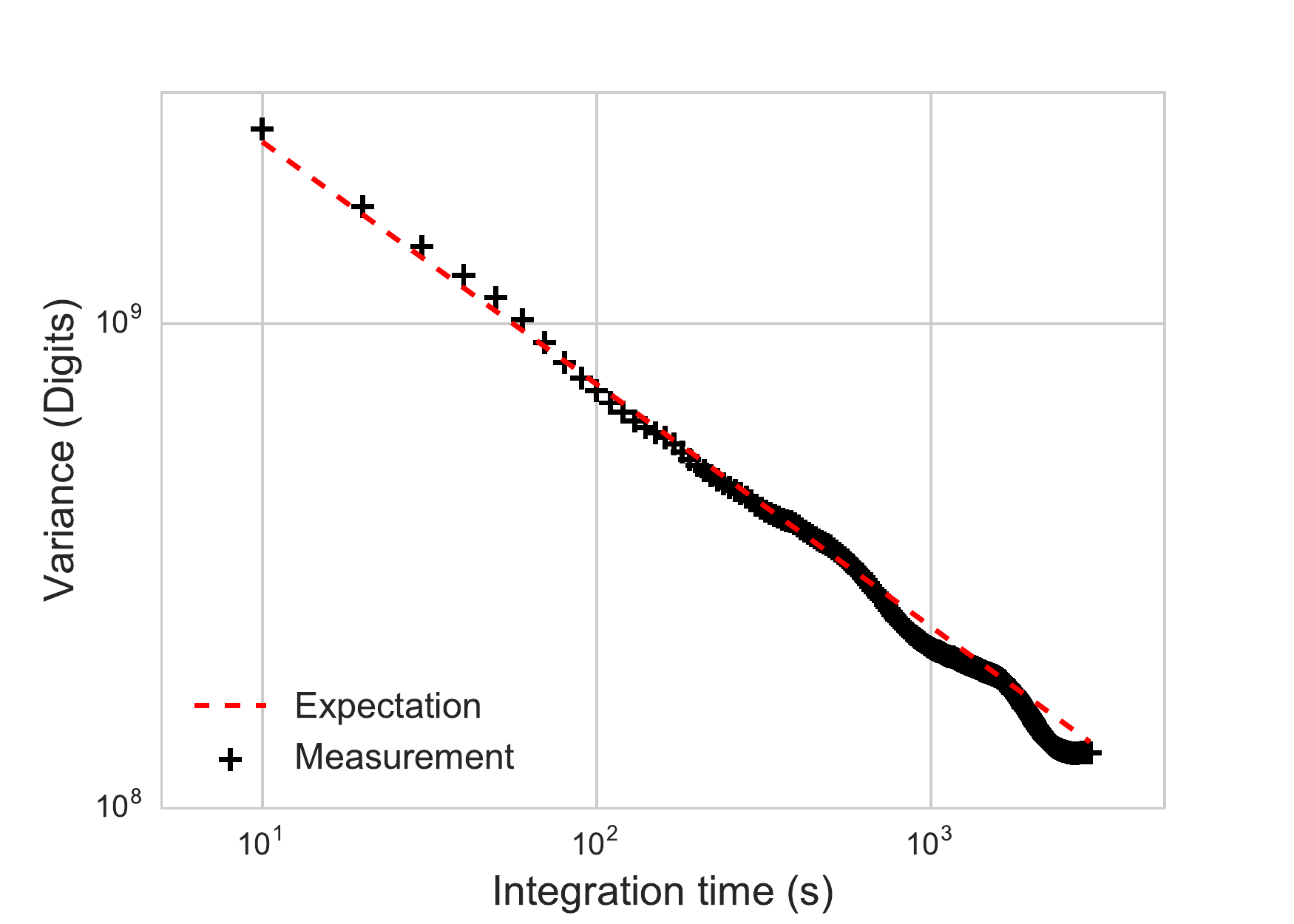}  
  \caption{PSD measurement from the FFT-spectrometer in total-power and phase-switch modes (left), and 
  the corresponding Allan time variance measurement for the phase-switch mode (right). The left panel is smoothed 
  by a Gaussian filter with standard deviation of 3 Hz. It illustrates 
  how the phase-switch implementation effectively reduces the $1/f$-noise in the data. The right panel shows that 
  the Allan time variance measurement confirms that the noise does average down as inverse integration time. }
  \label{fig:psd}
  \end{center}
\end{figure*}

\begin{table}
\begin{center}
\caption{Instrument specifications.}
\label{tab:instrumentspecs} 
\begin{tabular}{ll}   
\hline \hline
Instrument parameter & Value and unit \\
\hline
Observatory location & Graenichen, Switzerland \\
              & $47^{\circ}~20'~23''$ N, $8^{\circ}~6'~42''$ E\\ 
Observatory altitude     & 469 m \\
Telescope geometry & 7 m parabolic dish \\
Telescope mount limits & $45^{\circ}<$Az$<315^{\circ}$\\
                                     & $5.4^{\circ}<$Ele$<87^{\circ}$ \\
Frequency range full sensitivity & 990 MHz - 1260 MHz\\
Frequency range reduced sensitivity & 970 MHz - 1360 MHz\\
Receiver temperature & $\leq 80$ K\\
Allan-time & $\geq 5000~s$\\
Channel separation & 48.8 KHz \\
Radiometric bandwidth & 51.5 KHz \\
Integration time per phase & 3~s\\
ADC resolution & 12 bits\\
\# of channels in spectrometer & 16584 \\
ADC clock rate & 1600 MHz\\
ADC input voltage & 2 V maximum at $50\Omega$\\
Temperature control receivers & Tambient -8 K, +18 K\\
Cold reference noise & 40 K $\pm10~K$\\
Dynamic range & $\cong70$ dB\\
Power consumption per antenna & 110 W\\
Maximum rf input power & -70 dBm\\
\hline
\end{tabular}
\end{center}
\end{table}

\begin{figure*}
  \begin{center}
   \includegraphics[scale=0.43]{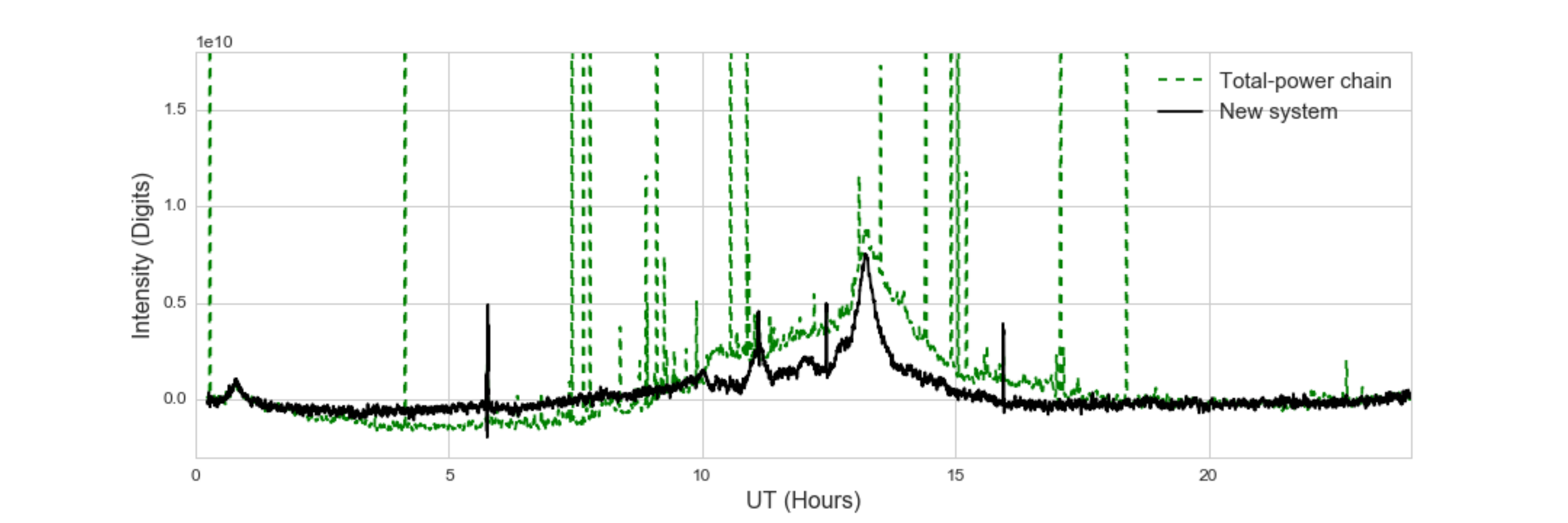}  
  \caption{Drift-scan data for one day at $\sim1008$ MHz. Data from the total-power chain (dashed green) is 
  overlaid with data from the new system (solid black) for comparison. Both data have been smoothed over 10 
  pixels ($\sim 1$ minute) and the median level over the whole day has been subtracted. One can clearly see 
  a less stable baseline in the total-power chain. The RFI level is also higher in the total-power chain, which is 
  mainly due to the fewer filters in the chain.}
  \label{fig:drift}
  \end{center}
\end{figure*}

\subsection{Laboratory measurements}
\label{sec:commission_lab}

Measurements in the laboratory were performed to test the entire electronic chain from the PCR to the 
recording of the FFT-spectrometer. The main focus at this stage is to test for the noise temperature 
of the PCR ($T_{rx}$) and the $1/f$-noise from the electronic chain. 

The PCR is expected to be the main contributing factor to the total noise temperature. As a result, an 
important laboratory measurement is the quantification of the noise temperature of the PCR when 
decoupled from the rest of the system (the sky, the horn, the cabling and the network).  
We perform the measurement by comparing the PCR recording under the known ambient temperature 
and liquid nitrogen temperature ($\sim78$ K). \Fref{fig:trx} shows the mean receiver noise temperature 
of the PCR over the full frequency range. We note that this noise temperature is a significant improvement 
over the total-power chain, which is about 200 K. This improvement mainly comes from the 
cancellation of $1/f$-noise from both the PCR and the phase-switch implementation. 
We also overlay in \Fref{fig:trx} the noise temperature measured for the entire system including the sky 
measured at the telescope, or the system temperature $T_{sys}$ (see \Aref{sec:sv} for the calculation of 
$T_{sys}$). We note that the $T_{sys}$ is very close to $T_{rx}$, and the difference is consistent with the 
temperature of the background sky. This demonstrates that other sources of noise in our 
system are small. Note that there are also potential errors due to the inaccuracy in the ambient temperature 
measurement and the liquid nitrogen temperature.  

To measure the $1/f$-noise in our laboratory data, we first look at the power spectrum density, or PSD, of the data. 
The PSD for each frequency channel is calculated by taking the Fourier transform of the time-sequenced data. 
The PSD measured in the laboratory gives a measure of the instrument $1/f$-noise -- pure white noise 
will have a flat power spectrum while pure $1/f$ (pink) noise will show a power spectrum with a slope of -1 
when plotting on a log-log scale. As both the PCR and the phase-switch implementation of the system are 
designed to reduce the $1/f$-noise, it is important to verify that the $1/f$-noise is indeed under control. The 
left panel of \Fref{fig:psd} shows the power spectrum from a single frequency channel in the FFT-spectrometer 
with and without implementing the phase-switch 
operation. We see that the $1/f$-noise is greatly reduced when operating in the phase-switch mode. This also 
demonstrates that with the PCR alone, we still have significant $1/f$-noise in the data. Another way to 
examine the residual $1/f$-noise in the data is to calculate the Allan time variance as a function of 
integration time \citep{Allan1966}. In the case of pure white noise, the Allan time variance is inversely 
proportional to the integration time. Here we follow the measurement technique proposed by 
\citet{Ossenkopf2008} and derive the Allan time variance curve of our system. The right panel of 
\Fref{fig:psd} shows the result of this measurement for a single frequency in phase-switch mode, where 
we recorded data for 18 hours. The plot shows clearly that the final Allan time variance of our electronic 
chain follows nicely the expectation for a white-noise system. This again suggests that the instrument is 
stable on time scales of at least tens of hours.

\begin{figure}
  \begin{center}
   \includegraphics[scale=0.43]{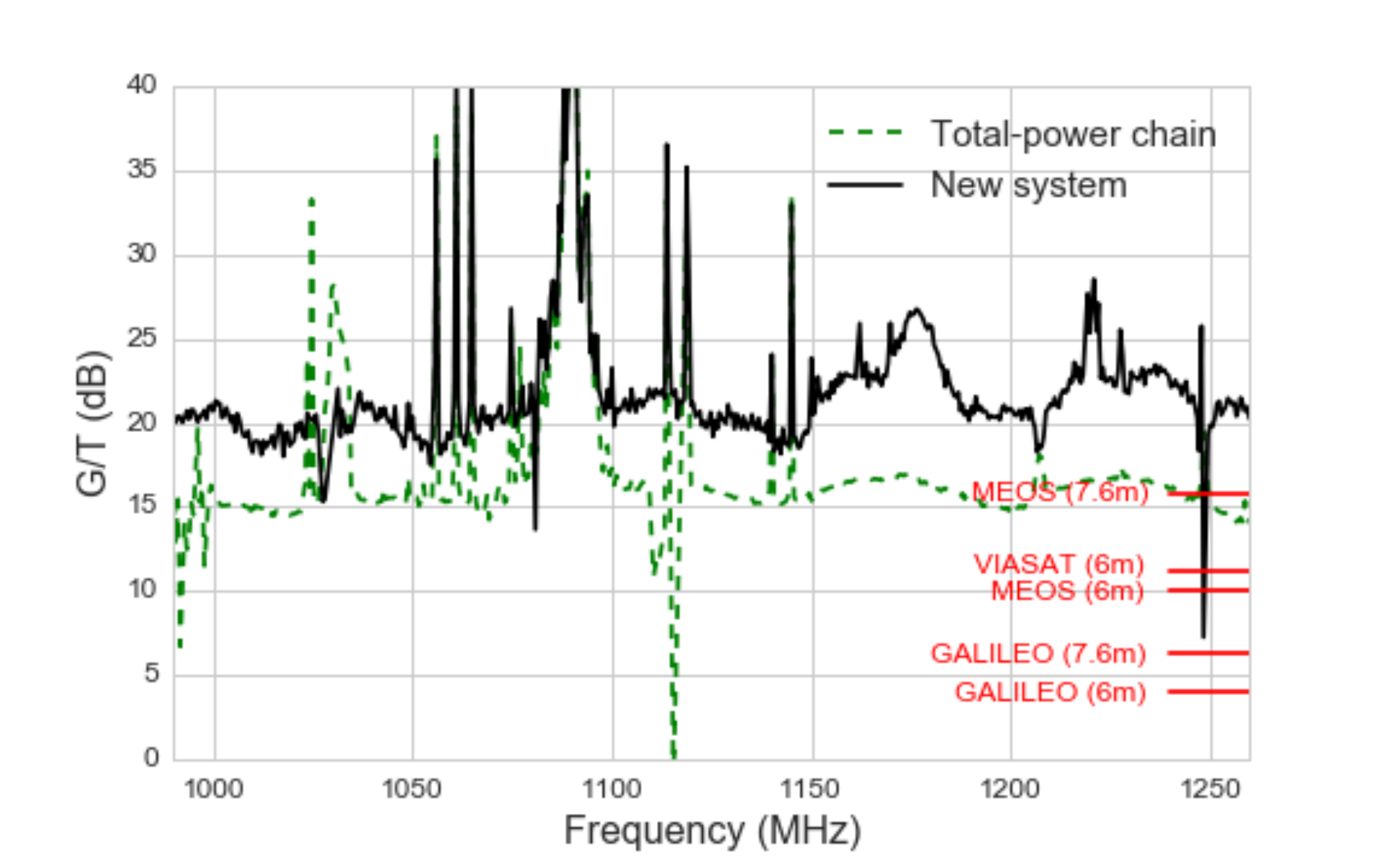}  
  \caption{$G/T$-factor calculated using Cassiopeia A, for the total-power chain (dashed green) and the new 
  system (solid black). The large peak at 1080 MHz is an artefact from the high-level emissions of the Automatic 
  Dependent Surveillance radars and altimeters. Also overlaid are some average $G/T$-factors for other dishes 
  of similar size.}
  \label{fig:gt}
  \end{center}
\end{figure}

\subsection{On-site measurements}
\label{sec:commission_site}
The full electronic chain was installed at Bleien on November 10, 2015 and the full system began operation 
soon after. The on-sky commission period continued till December 14, 2015, during which we performed 
various tests to examine and tune the performance of the new system. A set of general calibration tests during 
this period and in the following SV stage are described in \Aref{sec:sv}. Here we only present 
the qualitative comparison between the new system and the total-power chain to demonstrate the improvement 
in the data quality from the new innovative hardware design. 

\Fref{fig:drift} shows the time-sequenced data for the mean of 20 frequency channels around 1008 MHz for 
a characteristic day for both the total-power chain and the new system. The data plots shown have been 
median-subtracted and scaled to have the same signal strength for easy comparison. The telescope is 
parked at position Az=200.0$^{\circ}$ and Ele=46.0$^{\circ}$ at this date, which corresponds to mapping 
the sky at $DEC$=4.9$^{\circ}$. The telescope beam drifts through the Milky Way at $\sim13.5$ UT, which 
can be seen clearly on the plot. Both data sets are smoothed with a top-hat filter of 50 pixels. From the plot, 
we find that the baseline drift in the new system is much reduced compared to the total-power chain. We 
also see that this stability comes at the price of a higher noise level. We note that there are also 3 additional 
high- and low-pass filters in the new system compared to the total-power chain, which is the main reason for 
reduced RFI contamination in the new system.  

A useful measure of the instrumental performance is the ``$G/T$-factor'' \citep{Baars1977, Kildal1997} 
defined through 
\begin{equation}
\frac{G}{T} = \frac{8k\pi (Y-1)}{S \lambda^{2}}
\end{equation}
where $k$ is the Boltzmann constant, $Y$ is the Y-factor\footnote{The Y-factor is a measure of the 
ratio of the signal+background and the background and is commonly used in radio astronomy. It is 
defined as the ratio of the flux at the center of the source over the background.} 
for a source with known flux $S$, and $\lambda$ is the wavelength of interest. For a given source and 
wavelength, the $G/T$-factor scales like $Y-1$, which is the ratio of the signal to the background. The 
advantage of using this figure-of-merit is that it does not depend on the measurement of beam size, 
system temperature and other method-dependent procedures. The only data required is a 
transit measurement of a known source. For dishes of the same size, the $G/T$-factor is a measure 
of the performance of an instrument. \Fref{fig:gt} shows the $G/T$-factor calculated for the total-power chain 
and for the new system. We find that the design of the new system effectively increases the $G/T$-factor by 
more than 5dB $\approx$ 316\% in almost all frequencies. This again demonstrates the power of the PCR 
and the phase-switch implementation. We also overlay the $G/T$-factors for several other radio 
telescopes\footnote{These $G/T$ values are taken from the  International Telecommunication Union  (ITU) 
report Rec. ITU-R S.733-2: ``Determination of the $G/T$ Ratio for Earth Stations Operating in the Fixed 
Satellite Service''.} with dishes of similar sizes to show that the performance of our instrument is 
above average amongst similar-class telescopes.

\section{Conclusion}
\label{sec:conclusion}

In this paper we describe the design of an integrated system at the Bleien Observatory for 
mapping the Galaxy. 
The system is designed to map the Milky Way in the L-band frequency range 990--1260 MHz with a 
7m single-dish telescope. The ultimate science goal of this system is to provide a set of new data for 
studying the foreground effect of future low-redshift H$_{\rm I}$ intensity mapping cosmology experiments 
such as BINGO and HIRAX. Specifically, the data from this system will fill 
the gap between the Haslam map at 408 MHz and the various maps constructed at the 21 cm frequency 
(1420 MHz). This data will also allow for explorations in different areas of Galactic astronomy. We 
describe the hardware upgrade of the 7m single-dish telescope at the Bleien Observatory from 
a more conventional electronic chain to a high-performance, stable system. We also demonstrate that 
the improvement in the data quality is significant.

Several innovative designs in the hardware system result in a very stable system (Allan-time $>$ 5000 s) 
with high dynamic range (70 dB) and high frequency resolution (50 kHz), which is important for the 
large-scale Galactic map. These innovations include the pseudo correlation receiver and the cold 
reference source within the receiver, the FFT-spectrometer and the phase-switch operation of the 
whole system. Some of these components have been well tested and developed in the microwave 
instruments for cosmic microwave background (CMB) measurements, but 
for radio astronomy, these technologies driven by large-scale cosmology are relatively new. The 
hardware system presented in this work provides an environment to test many of these new ideas 
for a radio telescope. 

The hardware system described in this paper together with the software pipeline developed in the 
companion paper A16 can be fed into the planning of a generic Galactic mapping survey. Future potential 
upgrades of this system include building a parallel electronic chain for the second polarization, 
installing ground-shields to reduce ground-pickup and RFI, and making the FFT-spectrometer more 
programmable so that more powerful functionalities in the spectrometer can be used. 
One interesting extension is to use the internal correlation function in the FFT-spectrometer to replace 
the front end correlation receiver \citep{Kooi2004}. This potentially could save hardware cost significantly.

\section*{Acknowledgement}

We thank apprentice Manuela Wipf for manufacturing the phase switched pseudo correlation 
receiver, the heterodyne receiver and the instrument controller. We also thank the mechanical 
workshop of ETH for production of dozens of mechanical parts for the receiver chain. We 
thank farmer Andreas Brunner and Marc Furrer from Electricity Power Supply Gr\"{a}nichen for
technical support with their hydraulic lifting ramp during installation of the heavy corrugated 
horn-antenna.


\appendix

\section{Science Verification}
\label{sec:sv}
The Science Verification (SV) period took part from mid December 2015 to end of May, 2016 after 
commissioning. During the SV, observations were carried out to mimic a real survey in terms of the 
survey strategy and scheduling. 

Based on analyses of data from commissioning considerations of resolution, total coverage, 
signal-to-noise, calibration and turnaround, the scanning strategy for SV is composed of the following 
elements: 
\begin{itemize}
\item Each 16-day periods is composed of 2 ``calibration days'' and 14 ``survey days''. The calibration 
days are planned in the beginning and the middle of the season. This pattern is used for the entire survey.
\item The first part of the survey consists of seven 16-day periods, where 5 are done at Az=200$^{\circ}$ 
to fill in the south part of the sky, and the other two seasons are done at Az=280$^{\circ}$ and 
Az=314$^{\circ}$ to fill in the north part of the map.
\item The map from the data described above is then used for planning the second part of the survey, 
where the largest masked positions are re-observed. This second part took about 1 month.
\end{itemize}

Data was taken to mimic a large-scale Galactic survey, resulting in a final 
sky coverage of $\sim$29,393 deg$^{2}$ in the region $-33^{\circ}<$Dec$<60^{\circ}$. The SV 
observations included 139 survey days and 21 calibration days. The data was analyzed using 
the software developed in A16, and some preliminary are also shown therein. 

\section{Additional tests on the overall system}
\label{sec:sv_tests}

In this appendix we present a series of tests on the hardware system. 
These tests are important for the planning of a generic Galactic survey. All tests 
were performed during commissioning and the SV period.

\subsection{Site RFI characterisation}
\label{sec:rfi}

\begin{figure*}
  \begin{center}
   \includegraphics[scale=0.45]{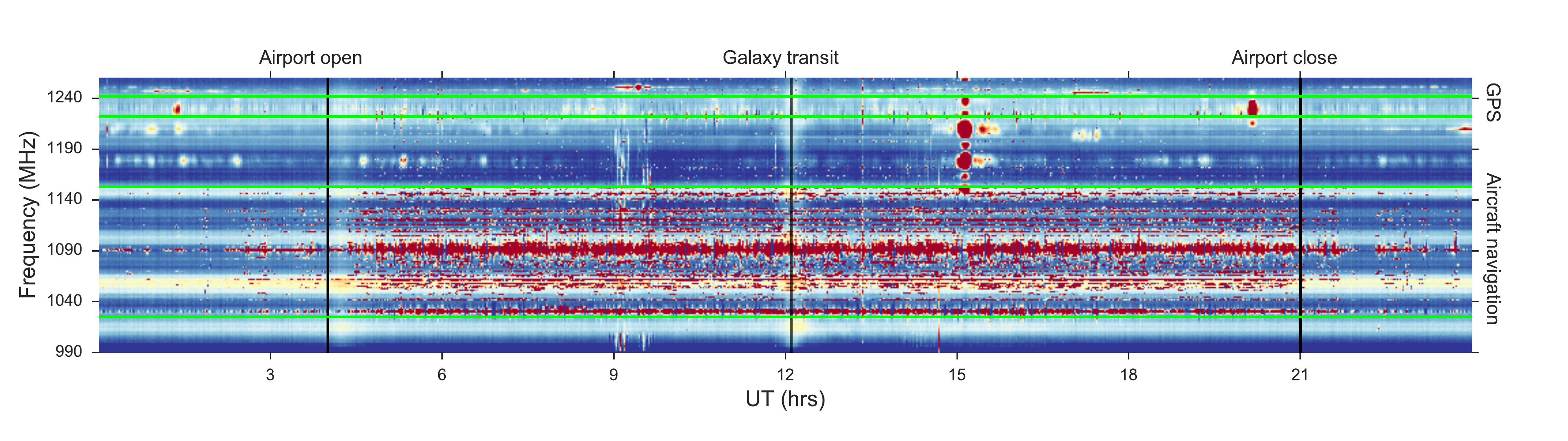}  
  \caption{Example of the raw time-frequency plane from the FFT spectrometer taken on May 10, 2016. 
  Specific bands in time and frequency are marked to show the most pronounced RFI sources, including the 
  aircraft navigation band and the GPS satellite band. Also marked is the expected crossing of the galaxy. }
  \label{fig:tod}
  \end{center}
\end{figure*}

\begin{figure*}
  \begin{center}
   \includegraphics[scale=0.5]{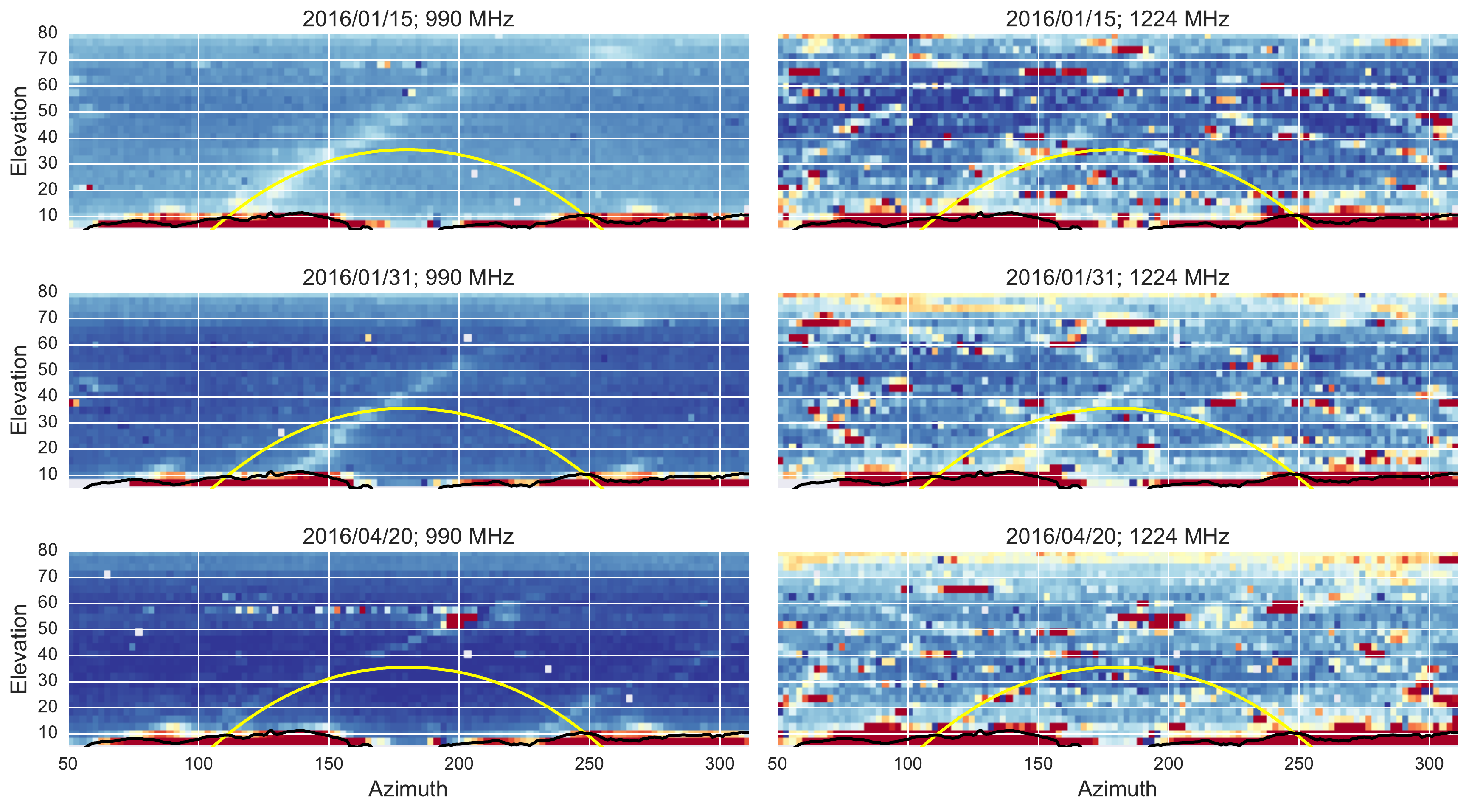}  
  \caption{Terrestrial scan of the observable region for the 7m dish at the Bleien Observatory for three 
  different days (January 15, January 31 and April 20 of 2016 from top to bottom). The left 
  panels show the maps for a relatively clean channel (990 MHz), whereas the right panel is more 
  contaminated by RFI (1224 MHz). The color scale is the same for panels of the same frequency 
  channel. The color scales for the left 
  and the right panels span the same range, but adjusted so that the background levels are similar for 
  easy visualization.}
  \label{fig:rfi}
  \end{center}
\end{figure*}

As mentioned in \Sref{sec:bleien}, one major challenges for our system is the RFI contamination at the Bleien 
Observatory. In addition to the site conditions, the frequency range of our instrument sits in the L-band (1-2 GHz) 
where many commercial, navigation, and surveillance applications lie. \Fref{fig:tod} shows one example 
of the raw time-frequency plane data from the FFT spectrometer and the most pronounced RFI sources.  
In order to understand the frequency range and potential locations of strong RFI emission 
sources, we perform three 2D terrestrial scans within the range of positions accessible by the telescope. 

All three scans were performed in identical patterns. Each scan moves through the positions 
$47^{\circ}<$Az$<313^{\circ}$ for 32 elevation angles equally spaced in the range $6^{\circ}<$Ele$<86^{\circ}$. 
The limits are imposed by the limits that the telescope can physically point to due to trees on the site, while 
the elevation interval is chosen to be just smaller than the beam size. Each horizontal scan takes $\sim20$ 
minutes.  

\Fref{fig:rfi} shows examples of the resulting terrestrial map generated from the data for the two frequency 
channels in the three scans. The plots are in terrestrial coordinates, Azimuth (Az) and Elevation (Ele), and are 
shown without spherical projection. The black line traces out the horizon from the mountains surrounding 
the area, which can be seen to emit thermal radiation much higher than the sky. Two regions of additional 
emission beyond the black line at Az=85$^{\circ}$ and 265$^{\circ}$ are known manmade structures (the 
control room of the telescope and the farmer's house). The yellow line traces the geo-stationary path, where 
there is a much higher chance of seeing strong point source emissions from the satellites. We note that the 
trajectories of other types of satellites on these plots will be more complicated. 

The left panels show a relatively clean frequency channel, where only few RFI pixels are seen. The 
galaxy can be seen clearly as a stripe across (from Az=115$^{\circ}$ at Ele=10$^{\circ}$ to Az=250$^{\circ}$ 
at Ele=75$^{\circ}$ for the first two scans and from Az=240$^{\circ}$ at Ele=10$^{\circ}$ to Az=300$^{\circ}$ 
at Ele=40$^{\circ}$ for the last scan), and there is smooth background gradient coming from the hot horizon 
leaking into the side- and back-lobe of the telescope. We also note that the baseline is not the same for all 
three days. The right panels show the maps at a frequency channel where there is strong RFI emission 
everywhere, channels like these are in general not usable, as the resulting maps will be heavily masked. 

Comparing maps from the same frequency channel across days, we observe that for the clean channel, 
(1) the position of the galaxy shifts over time (2) geostationary satellites do not change positions and (3) 
there are a few more point-like sources in the maps, which would be other types of satellites as the 
positions are not exactly the same in the two maps. For the contaminated channel, on the other hand, 
the emissions are scattered randomly across the field aside from the geostationary satellites.

These RFI maps help us to ensure that the drift-scan positions do not point into known high RFI regions 
during the scans. It also gives us a qualitative understanding of the RFI environment in the area. We find 
that within our frequency range of interest, there are a few localized clean frequency bands at the low 
frequency end and around 1180 MHz, but most other frequencies outside these bands are contaminated 
by RFI at a median to serious level. This again manifests the challenge in observing in L-band, where the cell 
phone (925 --960 MHz\footnote{Although this frequency range is not directly in our band, the very strong 
signal can saturate and leak into our band if the filters were not properly installed.}), aircraft navigation 
(1025--1150 MHz), and GPS satellite (1176 MHz, 1228 MHz) 
communication bands lie, and also highlights the importance of a good RFI mitigation algorithm. 

\begin{figure*}
  \begin{center}  
         \includegraphics[scale=0.43]{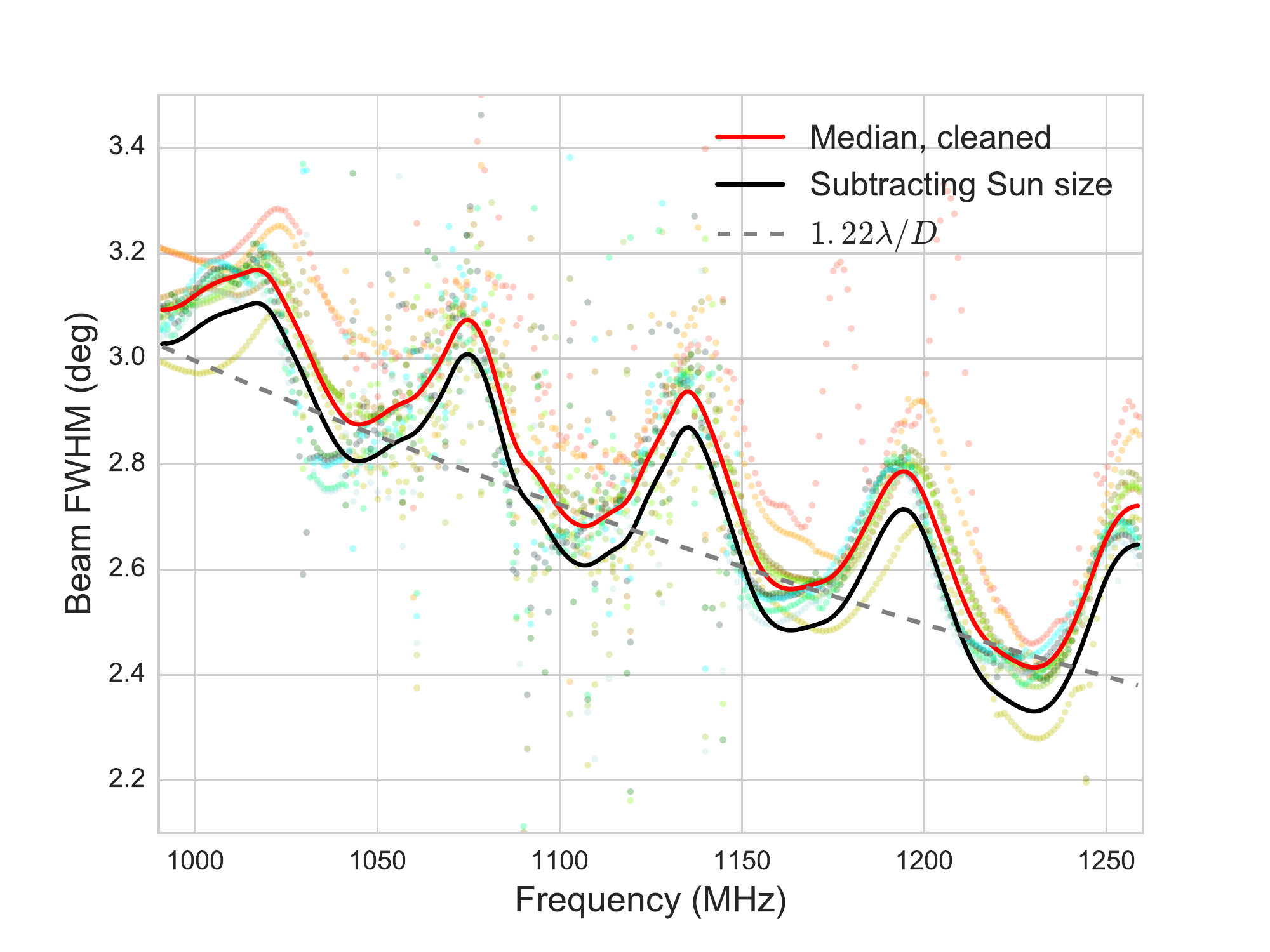} 
      \includegraphics[scale=0.43]{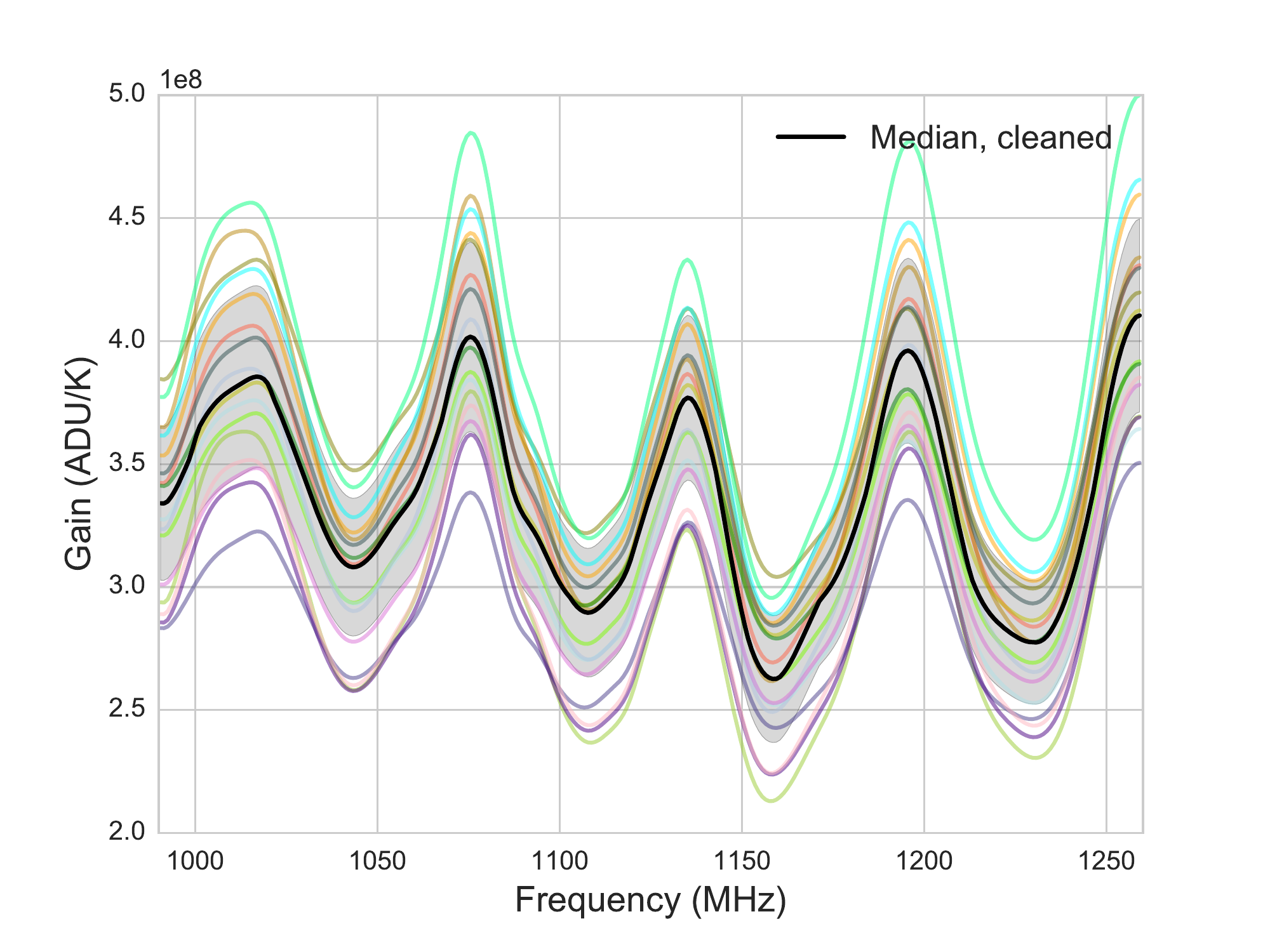} 
   \caption{Left: beam size measured using the Sun as a function of frequency. The data points of different 
   colors show the measurement performed on different days distributed over the Science Verification 
   period, while the median, cleaned beam size measurement is shown in red. Subtracting the Sun 
   size in quadrature gives the black line, which is our final beam size. Right: the gain calibration results 
   measured from Cassiopeia A as a function of frequency is shown. Different colors represent measurements 
   performed on different days. The grey shaded region shows the standard deviation over all the curves.}
  \label{fig:gain}
  \end{center}
\end{figure*}
\subsection{Pointing calibration}
\label{sec:pointing}

A general Galactic map needs to be built based on knowing where the telescope is pointing. Thus it is 
important to 
understand the uncertainty in the pointing of the telescope, or, the difference between the recorded 
telescope pointing coordinates and the actual pointing coordinates. We performing 2D raster scans 
of the Sun, a bright source whose position on the sky we know accurately. As we do not expect this 
uncertainty to change dramatically over time, we only perform two measurements before and during 
the survey. Also, since the FFT spectrometer does not have a sufficiently fine time resolution, we use 
another instrument (the Callisto spectrometer) to perform this measurement. The Callisto spectrometer 
is connected to the same horn antenna as the FFT spectrometer, but using a different electronic chain. 
This, however, should not affect the validity of the pointing measurement.  

We plot the scanned result on the coordinates according to the telescope recording, where the center is 
where we expect the Sun to be. If the pointing of the telescope was perfect, we would expect the center 
of the Sun to be measured at (0,0) in these coordinates. 
For the 3 measurements that was made, we find the 
mean and 1$\sigma$ uncertainty on the mean for the pointing errors over all frequencies to be 
$4.17\pm0.15$ arcmin ($-1.56\pm1.41$ in Az direction, $-3.87\pm1.68$ in Ele direction)
$11.56\pm0.21$ arcmin ($-11.56\pm1.60$ in Az direction, $-0.32\pm2.57$ in Ele direction) and 
$11.42\pm0.20$ arcmin ($-11.38\pm1.74$ in Az direction, $0.97\pm2.26$ in Ele direction). 

We note that two other effects need to be considered when interpreting the results. First, the 
telescope is constantly moving in the Az direction when performing these 2D scans, meaning that when 
the telescope is moving fast, the pointing in that direction will less accurate compared the other direction, 
and also less accurate compared to when the telescope is static (i.e. in a drift-scan mode). 
We indeed see this trend in the data. For the latter two measurements, the scanning velocity was much 
faster than the first one, and we see there is a much larger error in the Az direction. Second, there is an 
inherent inaccuracy in the telescope recording coming from the fact that it only records every $\sim7$ 
seconds. In a drift-scan scenario, this means that the pointing cannot be known better than $\sim5$ 
arcminutes, which is larger than the static pointing error we can infer from the Ele direction 
measurement discussed above, and the Az direction in the first measurement. 

To summarize, we expect the pointing error in our data during the entire survey to be limited by 
the telescope recording at $\sim5$ arcminutes.

\subsection{Beam size}
\label{sec:beam}
The most important aspects of the beam is the beam size, as it directly determines the limiting resolution 
of the telescope. We measure the beam size using 1D transit measurements of the Sun, as it provides 
high signal-to-noise for the measurement. For our telescope beam ($\sim2.7^{\circ}$ FWHM), the Sun is 
nearly a point source ($\sim0.63^{\circ}$ FWHM). Nevertheless we subtract a fixed size of the Sun in 
quadrature when quoting the measured size of the beam. We monitor the beam size throughout the entire 
Science Verification period. 

12 out of the 21 calibration days contain Sun measurements, and 2--4 measurements were made each 
day. We first fit a Gaussian profile to each transit measurement in each frequency channel and calculate 
the resulting FWHM of the profile as a function of frequency as shown in the left panel of \Fref{fig:gain}. 
The cleaned, median measurement and the beam size after removing the Sun contribution is overlaid in the 
figure together with the expected beam size for an idealized 7m dish. We observe several features in these 
results. First, on top of the expected $1.22 \lambda/D$ trend, there are so-called ``standing-wave'' features 
in the beam size measurements. This is similar to that reported in \cite{Briggs1997,Popping2008,Chang2015}, 
and is caused by the physical standing waves generated from the interference of the electromagnetic wave 
reflecting between the front end and the dish, as well as in the electronics. Second, the measurements 
are significantly noisier in the frequency range 1025--1150 MHz due to RFI contamination. Finally, there 
appears to be a $\sim10\%$ variation in the measured beam size over time. This can be due to the size 
variation of the observed Sun and the change in beam size over elevation.     

\subsection{Gain calibration and system temperature}
\label{sec:gain}
We demonstrate the flux calibration of our system based on Cassiopeia A, as its spectra is well-understood 
\citep[e.g.,][]{Baars1977}, and it is measured with a good signal-to-noise in our instrument.
In A16, we describe the procedure of the gain calibration as built in the analysis pipeline. The output of 
this procedure is the ``gain'' $G(\lambda)$ (Eq. C.2 in A16), or the conversion between the 
instrument-recorded values (ADU) and the physical surface temperature (K). In the 
left panel of \Fref{fig:gain} we show the gain calculated as a function of frequency from Cassiopeia A 
measurements taken throughout the Science Verification period. We again see the characteristic 
standing-wave patterns similar to that seen in the beam size. The variation over this time period is 
roughly at the 10\% level. 

We can then continue to estimate the on-sky system temperature by dividing the baseline level of 
the data (the smooth background without astronomical sources and RFI) by $G(\lambda)$. With the 
Cassiopeia A measurements shown in the right panel of \Fref{fig:gain}, we derive the median system 
temperature as shown in \Fref{fig:trx}. As discussed in A16, the system temperature fluctuates at the 
few K level and mainly comes from the temperature fluctuation in the horn and other exposed, 
non-regulated components. These will be improved upon in the future.

\end{document}